\documentclass[a4paper,11pt]{article}
\usepackage[utf8]{inputenc}
\usepackage{mathtools,amsmath,amssymb}
\usepackage{graphicx}
\usepackage{lmodern}
\usepackage[T1]{fontenc}
\usepackage{microtype}
\usepackage{hyperref}
\usepackage{xspace}
\usepackage{bibentry}
\usepackage{mcite}
\usepackage{cite}
\usepackage[left=1.7cm,right=1.7cm,bottom=2cm,top=2cm]{geometry}
\usepackage[usenames,dvipsnames,svgnames,table]{xcolor}
\definecolor{todocolor}{HTML}{D7E1E5}
\usepackage[textsize=tiny, 
            backgroundcolor=todocolor, 
            bordercolor=todocolor, 
            linecolor=todocolor, 
            textwidth=3.9cm]{todonotes}

\usepackage{comment}
\usepackage{rotating}

\newcommand{\NN}{\mathbf{N}}

\newcommand{\dmonQ}{{\rm U}}
\newcommand{\dmonT}{\mathcal{U}}

\newcommand{\qmon}{\mathrm{M}}

\newcommand{\osc}[1]{\boldsymbol{\mathrm{#1}}}

\newcommand{\oa}{\osc{a}}

\newcommand{\oad}{\osc{\bar{a}}}

\newcommand{\LL}{L}

\newcommand{\KQL}{\mathrm{K}}
\newcommand{\KQR}{\hat{\mathrm{K}}}

\newcommand{\ID}{\mathrm{I}}

\DeclareMathOperator{\tr}{tr}

\numberwithin{equation}{section}

\begin{document}

\begin{titlepage}
\hfill {\small\texttt{NORDITA 2022-055}}
 \vspace{0.1in}

\begin{center}
 \textbf{\Large Algebraic Bethe ansatz for Q-operators of\\[1mm]
 the open XXX Heisenberg chain with arbitrary spin}
\\\vspace{0.2in}
{\large Rouven Frassek$^{\,a}$ and István M. Szécsényi$^{\,b}$}
\\[0.2in]

$^{\,a}$ University of Modena and Reggio Emilia, FIM,\\
Via G. Campi 213/b, 41125 Modena, Italy
\\[0.1in]
$^{\,b}$ Nordita,  Stockholm University and KTH Royal Institute of Technology,\\  Hannes Alfvéns väg 12,  SE-106  91  Stockholm,  Sweden
\\[0.6in]
 \end{center}
 \vspace{-.2in}
\begin{center}
\textbf{\large Abstract}
\end{center}
\begin{center}
\begin{minipage}{400pt}
 \noindent  
 In this note we construct Q-operators for  the spin s  open Heisenberg XXX  chain with diagonal boundaries   in the framework of the quantum inverse scattering method.
Following the algebraic Bethe ansatz we  diagonalise  the introduced  Q-operators using the fundamental commutation relations.  
By  acting on Bethe off-shell states and  explicitly evaluating the trace in the auxiliary space we compute the eigenvalues of the Q-operators in terms of Bethe roots  and show that the unwanted terms vanish if the Bethe equations are satisfied. 
\end{minipage}
\end{center}
\vspace{0.3in}
{\small
\tableofcontents
}
\end{titlepage}

\newpage
\section{Introduction}\label{sec:intro}
In 1931  Hans Bethe solved  the XXX Heisenberg spin chain with cyclic  boundary conditions using the coordinate Bethe ansatz \cite{Bethe1931ZurTD}. A few decades later, the method was generalised to   the XXX Heisenberg spin chain open boundary conditions \cite{gaudin1983fonction,Alcaraz_1987}. The coordinate Bethe ansatz was then formulated in the form of the algebraic Bethe ansatz \cite{Faddeev:1996iy} where the case of open chains was introduced in \cite{Sklyanin1988} following ideas of \cite{Cherednik:1984vvp}. It relies on the construction of the transfer matrix within the quantum inverse scattering method. The spectrum then follows from Baxter's TQ-equation and is expressed in terms of Baxter Q-functions \cite{Baxter2007}.  
In the last decade, the theory of Q-operators whose eigenvalues yield the Baxter Q-functions and the study of their functional relations (QQ-relations) for closed spin chains was   developed  from various different perspectives of theoretical physics. Besides the developments arising from the study of the representation theory of quantum groups and solutions of the Yang-Baxter equation, see in particular \cite{Boos:2012tf,Frenkel:2013uda,Mangazeev:2014bqa,Meneghelli:2015sra,Boos:2017mqq,Braverman:2016pwk,Frassek:2020lky,Frassek:2020nki,Ferrando:2020vzk,Tsuboi:2021xzl,Frassek:2021sdm,Frassek:2021ogy,Miao:2020irt}, important contributions came from the ODE/IM correspondence \cite{Dorey:2007zx} and more recently from 4d Chern-Simons theory \cite{Costello:2017dso,Costello:2018gyb}. The majority of the results however concerns spin chains with periodic boundary conditions.

Open spin chains are  more difficult to study as apart from solutions to the Yang-Baxter equation also solutions to the boundary Yang-Baxter equations are needed in Sklyanins construction of transfer matrices within the quantum inverse scattering framework \cite{Sklyanin1988}. These type of spin chains play an important role in condensed matter physics, stochastic particle processes but also appeared in the AdS/CFT correspondence. The first step towards understanding the Q-operator construction for these types of systems has been done by the authors in \cite{Frassek:2015mra} where the case of diagonal boundaries and spin $\frac{1}{2}$ in the physical space has been studied. The authors followed the ideas of \cite{Bazhanov1999,Antonov:1996ag,Rossi2002,Bazhanov2010a} and incorporated new degenerate solutions to the boundary Yang-Baxter equation in the construction of Q-operators.  The approach was generalised to the q-deformed case in the works \cite{Baseilhac:2017hoz,Vlaar}, see also \cite{Tsuboi:2018gfd} for some partial results of the higher rank case.
We further like to mention that, at least in the rational case, the  restriction to diagonal boundary terms is less restricted than it may seem. In fact it is known that at least for  the spectrum for a large class of models with non-diagonal boundaries coincides with the one of the diagonal boundaries, see~\cite{Melo:2004erd}. The eigenstates of these models are related by a non-local  similarity  transformation, cf. \cite{Alcaraz:1992zc,Frassek:2019imp}.  One of most famous examples of such chain appear in the study of the  symmetric exclusion process whose Markov generator can be mapped to the Hamiltonian of a spin chain with the aforementioned properties.

In this article we study the generalisation of  the construction of Q-operators to higher spin in the quantum space for which only partial results exist.  More precisely, for the q-deformed case universal monodromy operators were given in an appendix of  \cite{Baseilhac:2017hoz}.  The TQ relation for these universal T- and Q-operators of the XXZ spin chain has been given in  \cite{Tsuboi:2020uoh}.  The evaluation of the Q-operators that appeared in that work for finite spin $s$ in the rational limit  should yields the Q-operators defined here but the convergence of the trace in the auxiliary space has not been discussed explicitly in \cite{Tsuboi:2020uoh,Baseilhac:2017hoz}. 
The result that we show in here  is that the eigenvalues of the Q-operator of the open spin $s$ XXX Heisenberg chain with diagonal boundaries defined through the trace construction within the quantum inverse scattering method are  Baxter polynomials such that the action on an off-shell state reads
\begin{equation}\label{eq:result}
 Q(x)|\psi_m\rangle= \frac{1}{2m-p-q-2Ns}\prod_{k=1}^m(x-x_k)(x+x_k+1)|\psi_m\rangle+\text{unwanted terms}\,.
\end{equation} 
Here $m$ is number of Bethe roots $x_i$, $x$ is the spectral parameter, $p$ and $q$ are the boundary parameters, and $N$ is the length of the spin chain.
This relation is derived using the fundamental commutation relations following the works \cite{Korff:2004ev,Frassek:2015qra}  in analogy to the ordinary Bethe ansatz for the transfer matrix \cite{Faddeev2007}. Furthermore, in addition to computing the eigenvalue of the Q-operator (wanted term) we show that the unwanted terms vanish if the Bethe equations are satisfied. 
Since by construction the Q-operator commutes with the transfer matrix one may ask whether it is not sufficient to just compute the wanted term. The answer is negative as the Q-operator may also contain Jordan blocks and thus could also be  not diagonalisable. For an operators to be diagonalisable, one must show that it is a normal matrix, i.e. it commutes with its conjugate transpose. Instead of following this path, we show the vanishing of the unwanted terms, and that the Q-operator acts diagonally on the Bethe states. For some recent related studies in the case of the closed chain with inhomogeneities (that can straightforwardly be introduced into our computations) we refer the reader to \cite{Chernyak:2020lgw,Ryan:2020rfk}.

The paper is organised as follows: In Section~\ref{sec:openxxx} we remind the reader of the construction of the transfer matrix of the open spin $s$ XXX Heisenberg chain that is relevant for the Bethe ansatz which we briefly outline. Furthermore we provide the definition of the Q-operators. Section~\ref{sec:act} is devoted to the fundamental commutation relations relevant for us and contains the exchange relations of the Q-operators with the B-operators that will be relevant to compute the action of the Q-operators on the Bethe off-shell states which is given in Section~\ref{sec:Qonbethestate}. In the same section we compute the wanted term and show that the unwanted terms vanish. To improve the readability of the article we present the details of this latter computation  in an appendix. Finally, Section~\ref{sec:conc} contains the conclusion.

\section{Open XXX Heisenberg chain}
\label{sec:openxxx}
In this section we construct the relevant transfer matrices and Q-operators within the framework of the quantum inverse scattering method for open spin chains \cite{Sklyanin1988}. We will first introduce the transfer matrix used for the ordinary Bethe ansatz with spin $s$ representations at each site of the quantum space and spin $s=\frac{1}{2}$ in the auxiliary space in Section~\ref{sec:tmat}. Section~\ref{sec:bae} is devoted to the algebraic Bethe ansatz that diagonalises the introduced transfer matrix. The Q-operators are then defined in Section~\ref{sec:qop}.

\subsection{Transfer matrix}
\label{sec:tmat}
We begin with introducing the R-matrix 
of the XXX spin chain that intertwines two spin $\frac{1}{2}$ representations along with the spectral parameter $x\in \mathbb{C}$. The R-matrix reads
\begin{equation}\label{rmatrix2}
 {R}(x)=\left(\begin{array}{cccc}
              x+1&0&0&0\\
              0&x&1&0\\
              0&1&x&0\\
              0&0&0&x+1
             \end{array}
             \right)\,.
\end{equation} 
The R-matrix is used to define the boundary Yang-Baxter equations relevant for the open spin $s$ chain. For the double-row monordromy $\mathcal{U}$ that in our conventions contains the right boundaries the it takes the form
\begin{equation}\label{eq:BYBE-A}
R(x-y)(\dmonT(x)\otimes \ID)R(x+y)(\ID\otimes \dmonT(y))  = (\ID\otimes \dmonT(y))R(x+y)(\dmonT(x)\otimes \ID)R(x-y)\,,
\end{equation}
where $x,y\in \mathbb{C}$.
The double-row monodromy $\dmonT$ is a $2\times 2$ matrix with operatorial entries in the quantum space $V$, such that 
\begin{equation}
 \dmonT(x):\mathbb{C}^2\otimes V\to\mathbb{C}^2\otimes V\,.
\end{equation} 
The R-matrix acts non-trivially on $\mathbb{C}^2\otimes \mathbb{C}^2$.
It is convenient to denote the matrix elemens elements in the auxiliary space of $\dmonT$  by
\begin{equation}\label{eq:abcd}
 \mathcal{U}(x)=\left(\begin{array}{cc}
                       A(x)&B(x)\\
                       C(x)&D(x)
                      \end{array}\right)\,,
\end{equation} 
where the entries $A,B,C,D$ solely act on the quantum space  $V$. The 
transfer matrix is then definded by the trace in the $2\times 2$ auxiliary space after incorporating the K-matrix $\mathcal{K}$ for the left boundary
\begin{equation}\label{eq:transm}
 T(x)=\tr \mathcal{K}(x)\mathcal{U}(x)\,.
\end{equation} 
Here $\mathcal{K}$ is a $2\times 2$ matrix that acts triviallly on the quantumm space. It  obeys the boundary Yang-Baxter equation
\begin{equation}\label{eq:BYBE-B}
R(y-x)(\mathcal{K}(x)\otimes \ID)R(-x-y-2)(\ID\otimes \mathcal{K}(y))  = (\ID\otimes\mathcal{K}(y))R(-x-y-2)(\mathcal{K}(x)\otimes \ID)R(y-x)\,.
\end{equation}
The boundary Yang-Baxter equations \eqref{eq:BYBE-A} and \eqref{eq:BYBE-B}   ensure the commutativity of the transfer matrix \eqref{eq:transm} at different values of the spectral parameters $[T(x),T(y)]=0$.

It is well known that a solution to the boundary Yang-Baxter equation \eqref{eq:BYBE-A} can be constructed from the Lax matrix 
\begin{equation}
 \mathcal{L}(x)=\left(\begin{array}{cc}
                       x+\frac{1}{2}+S_3&S_-\\
                       S_+&x+\frac{1}{2} -S_3
                      \end{array}\right)\,,
\end{equation} 
with the $sl(2)$ generators satisfying the standard commutation relations 
\begin{equation}\label{eq:gens}
 [S_3,S_\pm]=\pm S_\pm\,,\qquad [S_+,S_-]=2S_3\,.
\end{equation} 
More precisely, the double-row monodromy can be writte in terms of single-row monodromies via
\begin{equation}\label{eq:drow}
 \mathcal{U}(x)=\mathcal{M}(x)\hat{\mathcal{K}}(x)\hat{\mathcal{M}}(x)\,,
\end{equation} 
where the single row monodromies are defined via
\begin{equation}
 \mathcal{M}(x)=\mathcal{L}^{[1]}(x)\mathcal{L}^{[2]}(x)\cdots \mathcal{L}^{[N]}(x)\,,\qquad  \hat{\mathcal{M}}(x)=\mathcal{L}^{[N]}(x)\mathcal{L}^{[N-1]}(x)\cdots \mathcal{L}^{[1]}(x)\,,
\end{equation} 
where the integer in the brackets $[k]$ with $k=1,2,\ldots,N$ denotes the spin chain site on which the Lax matrix acts non-trivially.
The K-matrix is chosen to be diagonal  and is itself a solution to the boundary Yang-Baxter relation \eqref{eq:BYBE-A} as imposed by \eqref{eq:drow} for $N=0$. The K-matrix for the right boundary  reads
 \begin{equation}
\hat{\mathcal{K}}(x)=\left(\begin{array}{cc}
                       q+x&0\\
                       0&q-x
                      \end{array}\right)\,.
\end{equation} 
Here $q\in\mathbb{C}$ denotes a boundary parameter.  
The other K-matrix on the left boundary is chosen to be of the diagonal form 
 \begin{equation}
{\mathcal{K}}(x)=\left(\begin{array}{cc}
                       p+x+1&0\\
                       0&p-x-1
                      \end{array}\right)\,,
\end{equation} 
with  $p\in\mathbb{C}$. For later purposes we note that the Lax matrices obey the unitarity relations
\begin{equation}\label{eq:unitarity}
 \mathcal{L}(x)\mathcal{L}(-x)=-\left(x-\frac{1}{2}\right)\left(x-\frac{1}{2}\right)\ID+C
\end{equation} 
where $C=\frac{1}{2}\left(S_+S_-+S_-S_+\right)+S_3S_3$.
In the following we will focus on finite-dimensional representations in the quantum space with $2s\in \mathbb{N}$ such that $\dim V=2s+1$ and $C=s(s+1)$.
For spin $s=\frac{1}{2}$ the generators in \eqref{eq:gens} are realised by the Pauli matrices $S_\pm=\sigma_\pm$ and $S_3=\sigma_3/2$.

\subsection{Bethe ansatz for the transfer matrix}\label{sec:bae}
In this section we review the Bethe ansatz for the transfer matrix \eqref{eq:transm} for arbitrary spin $s$ representations in the quantum space, see~\cite{Sklyanin1988}. As noted by Sklyanin, it is convenient to define the linear combination 
\begin{equation}\label{eq:tildeD}
\tilde{D}(x)=  D(x)-\frac{1}{1+2x}A(x)\,,
\end{equation}
of $A$ and $D$ operators in \eqref{eq:abcd} such that the transfer matrix can be written as 
\begin{equation}\label{AplusD}
    T(x)
    =2\frac{1+x}{1+2x}(p+x)A(x)+(p-x-1)\tilde D(x)\,.
\end{equation}
The Bethe ansatz then requires a reference state $|\Omega\rangle$ such that  
\begin{equation}\label{eq:actref}
    A(x)|\Omega\rangle=\alpha(x)|\Omega\rangle \,,\qquad     \tilde D(x)|\Omega\rangle= \tilde \delta(x)|\Omega\rangle \,,\qquad     C(x)|\Omega\rangle=0 \,.
\end{equation}
Here $A$ and $\tilde D$ act diagonally on the reference state. The explicit form of the eigenvalues  $\alpha$ and $\delta$ is not needed here but given in \eqref{eq:ADvac} for completeness. 
The Bethe off-shell states are then introduced via
\begin{equation}\label{eq:offshell}
 \left|I\right\rangle =
B(x_1)B(x_2)\dots B(x_m)\left|\Omega\right\rangle  \,,
\end{equation}
where   $I=(1,2,\ldots,m)$ denotes the ordered
set and $m$ is called the magnon number. The ordering is not important in this context as the operators $B(x_i)$ commute among themselves. 
The action of the transfer matrix on such Bethe off-shell state \eqref{eq:offshell} is given in terms of Baxter's TQ-equation plus some unwanted terms. More precicely using the relations \eqref{eq:acta} and \eqref{eq:actd}   we find 
\begin{equation}\label{eq:tq}
\begin{split}
  T(x)|I\rangle &=t(x)|I\rangle+2(x+1)\sum_{j\in I}\frac{G_j^I}{(x-x_j)(x+x_j+1)}B(x)\left|I\setminus\left\{ j\right\}  \right\rangle\,.
  \end{split}
\end{equation}
The wanted term $t(x)$, i.e. the eigenvalues of the transfer matrix are parametrised in terms of the Bethe roots $x_k$ with $k=1,\ldots,m$ via
\begin{equation}\label{eq:TQreal}
\begin{split}
 t(x) &=2\frac{1+x}{1+2x}(p+x)  \alpha(x)\prod_{k=1}^m\frac{(x+x_k)(x-x_k-1)}{(x-x_k)(x+x_k+1)}+(p-x-1){\tilde \delta}(x)\prod_{k=1}^m\frac{(x-x_k+1)(x+x_k+2)}{(x-x_k)(x+x_k+1)}\,.
  \end{split}
\end{equation}  
The unwanted terms in \eqref{eq:tq} that are proportional to $|I\setminus\{j\}\rangle$ are given by 
\begin{equation}\label{eq:GG}
G_j^I=(p+x_j){\mathcal{A}}^{I }_j-(p-x_j-1)\tilde {\mathcal{D}}_j^{I}
\end{equation} 
with the  coefficients 
\begin{equation}\label{eq:adcof}
 {\mathcal{A}}_{k}^{I}=  \frac{2x_k\alpha(x_k)}{1+2x_k}\prod_{j\in I\setminus\{k\}}\frac{(x_k+x_j)(x_k-x_j-1)}{(x_k-x_j)(x_k+x_j+1)}\,,\qquad 
\tilde{\mathcal{D}}_{k}^{I} =  \tilde{\delta}(x)\prod_{j\in I\setminus \{k\}}\frac{(x_k-x_j+1)(x_k+x_j+2)}{(x_k-x_j)(x_k+x_j+1)}\,.
\end{equation} 
The condition that the unwanted terms vanish then yield the Bethe equations that we compactly write as $G_j^I=0$ for all  $j\in I$.

We note that Q-functions can be introduced into the TQ-equation when expressing the products rational functions in the spectral parameter with the eigenvalues of the Q-operator in \eqref{eq:result}. In the case of $s=\frac{1}{2}$ we have verfyfied  this relation on the operatorial level to conclude that the operators constructed in \cite{Frassek:2015mra} are indeed Q-operators and in addition showed that the trace converges. Here we do not follow this path but directly diagonalise the Q-operators using the algebraic Bethe ansatz.

\subsection{Q-operators}\label{sec:qop}

Similar to the construction of the transfer matrix \eqref{eq:transm}, the Q-operators are built as the  trace over a K-matrix and a double-row monodromy 
\begin{equation}\label{eq:qop}
Q(x)=\tr \KQL  (x) U(x)\,.
\end{equation} 
However, as discussed below, here the trace is taken over an infinite-dimensional oscillator space.
The commutatitity of the transfer matrix \eqref{eq:transm} and Q-operators \eqref{eq:qop} is based on the boundary Yang-Baxter equations
\begin{equation}
L(x-y)\dmonQ_{}(x)L(x+y)\dmonT(y) = \dmonT(y)L(x+y)\dmonQ_{}(x)L(x-y)\,,\label{eq:BYBE-1}
\end{equation}
and
\begin{equation}
\bar L(y-x)\KQL_{}(x)\bar L(-x-y-2)\mathcal{K}(y) = \mathcal{K}(y)\bar L(-x-y-2)\KQL_{}(x)\bar L(y-x)\,.\label{eq:BYBE-2}
\end{equation}
Here the intertwining Lax matrices are
\begin{equation}\label{eq:plax3}
L(x)=\left(\begin{array}{cc}
              1&\oad\\
	      \oa&x+1+\NN
             \end{array}\right)\,,\quad\quad  \bar L(x)=\left(\begin{array}{cc}
              x-\NN&\oad\\
	      \oa&-1
             \end{array}
           \right )\,,
\end{equation}
see~\cite{Frassek:2015mra}. 
The operators $(\oa,\oad)$ satisfy the Heisenberg algebra
\begin{equation}\label{eq:heis} 
 [\oa,\oad]=1\,.
\end{equation}  
Furthermore we defined the number operator $\NN=\oad\oa$.  To generalisation the construction of $s=\frac{1}{2}$ to arbitrary spin we  employ the Lax operators 
\begin{equation}\label{rop1}
\mathcal{R}(x)=e^{\oa S_-}\,\frac{\Gamma(x+\tfrac{1}{2}-S_3)}{\Gamma(x+\tfrac{1}{2}-s)}\,e^{\oad S_+}\,,
\end{equation}  
found in \cite{Frassek2011} to built the monodromies. 
The oscillator algebra and the $\mathfrak{sl}(2)$ generators obeying the commutation relations given in \eqref{eq:heis} and \eqref{eq:gens}, respectively. For a given finite-dimensional  representation the operator \eqref{rop1} can be evaluated as a finite-dimensional matrix.
For example the Lax matrix for spin $1$ is of the form
\begin{equation}
\mathcal{R}^{s=1}\left(x-\tfrac{1}{2}\right)=\left(\begin{array}{ccc}
                    1&\sqrt{2}\oad&\oad^2\\
                    \sqrt{2}\oa&x+2\NN+1&\sqrt{2}\oad(x+\NN+1)\\
                    \oa^2&\sqrt{2}(x+\NN+1)\oa&(x+\NN+1)(x+\NN)
                   \end{array}
\right)\,.
\end{equation} 
For spin $s=1/2$ we get $L(x)=\mathcal{R}^{s=1/2}\left(x\right)$.
The auxiliary space and so the K-matrices in \eqref{eq:bop} remain unchanged.  Thus in analogy to \cite{Frassek:2015mra} the Q-operator is defined through \eqref{eq:qop} with the single-row monodromies
\begin{equation}\label{eq:qmons2}
 \qmon(z)=\mathcal{R}^{[1]}(z)\mathcal{R}^{[2]}(z)\cdots \mathcal{R}^{[N]}(z)\,,\qquad  \hat\qmon(z)=\mathcal{R}^{[N]}(z)\mathcal{R}^{[\LL-1]}(z)\cdots \mathcal{R}^{[1]}(z)\,.
\end{equation}  
that are used to build the double-row monodromy
\begin{equation}
 U(x)=\qmon(x) \KQR (x)\hat\qmon(x)\,.
\end{equation} 
The boundary K-matrices were given in \cite{Frassek:2015mra} and  do not depend on the quantum space and only depend on the oscillators. They read
\begin{equation}\label{eq:bop}
 \KQL(x)=\frac{\Gamma(-p-x-1-\NN)}{\Gamma(- p-x)} \,,\quad\quad  \KQR(x)=\frac{\Gamma( q-x)}{\Gamma( q-x-\NN)}\,.
\end{equation} 
The trace in \eqref{eq:qop} is taken over the infinite dimensional oscillator space $\tr X=\sum_{n=0}^\infty \langle n|X|n\rangle$ with lowest weight state $|0\rangle$ such that $\oa|0\rangle =0$.
The operator constructed in this way commutes with the transfer matrix \eqref{eq:transm}.  In addition to the boundary Yang-Baxter equation the verification of the commuativity boils down to the verification of the Yang-Baxter relations 
\begin{equation}
 \mathcal{L}(y)\mathcal{R}(x)L(x-y)=L(x-y)\mathcal{R}(x)\mathcal{L}(y)\,,
\end{equation} 
which has been shown in \cite{Frassek2011}. Here we make use of the unitarity relation \eqref{eq:unitarity} to relate the Yang-Baxter equations that are relevant, cf.~\cite{Frassek:2015mra}.

As discussed in \cite{Frassek:2015mra} there is a spin-flip symmetry that generalises to the spin $s$ case. By conjugating the boundary Yang-Baxter equations \eqref{eq:BYBE-1} and \eqref{eq:BYBE-2} with the spin-flip matrix $\sigma_1$ in the auxiliary space we find that we obtain another Q-operator  $Q_-$
when substituting
\begin{equation}\label{eq:spinflip}
 p\to -p\,,\qquad q\to-q\,,\qquad S_\pm\to S_\mp\,,\qquad S_3\to -S_3\,.
\end{equation} 
By construction this operator commutes with the transfer matrix.
The transformation \eqref{eq:spinflip} on the $su(2)$ generators can be written as a similarity transformation for finite-dimensional representations  and thus the Bethe ansatz we are going to present  works the same way on the other reference state. The Bethe ansatz on the same reference state is relevant for non-compact highest-weight  representation and has been discussed to some extend in the closed case in \cite{Korff:2004ev,Frassek:2015qra}. As discussed in \cite{Frassek:2017bfz} for the closed chain, the  Lax operators remain the same. Here we do not persue this further.
For $s=\frac{1}{2}$ the Q-operator $Q$ denotes the Q-operator $Q_+$ of \cite{Frassek:2015mra}. The two Q-operators are related by the QQ-relations that can be obtained by equating the TQ-equations for $Q_+$ and $Q_-$, see \cite{Nepomechie:2019gqt}.

\section{Action of the  Q-operator on B-operators}
\label{sec:act}
In this section we  derive the action of the Q-operator  \eqref{eq:qop} on a product of B-operators using the fundamental commutation relations that arise from the boundary Yang-Baxter relation \eqref{eq:BYBE-1}, see Section~\ref{app:fcrQ}.
 For this it is convenient to define auxiliary operators
\begin{equation}\label{eq:general_Q}
W_{i,j}(x) =  \tr  \frac{\Gamma(-p-x+i-\NN)}{\Gamma(-p-x+i)}\,\bar{\mathbf{a}}^{j}\left( \dmonQ(x)+\bar{\mathbf{a}}\dmonQ(x)\mathbf{a}\right)\,,
\end{equation} 
 that reduce to  the Q-operator for $i,j=0$ such that $W_{0,0}(x)=Q(x)$, cf.~Section~\ref{sec:auxop}.  The auxiliary operators $W_{i,j}$ obey the exchange relations 
\begin{equation}
 \begin{split}  
W_{i,j}(x)B(y) &=  (x-y)(x+y+1)B(y)W_{i+2,j}(x)+X_{i,j}(x,y)\,,\label{eq:QB_commutator_master}
 \end{split}
\end{equation}  
where
\begin{equation}\label{eq:X_definition}
\begin{split}
X_{i,j}(x,y) &=  -W_{i+1,j+1}(x)\left( (p+y-i)\frac{2y}{1+2y}A(y)-(p-y-1-i)\tilde{D}(y)\right) \\
&\quad\;+(p+y-i-1) (p-y-2-i)W_{i+2,j+2}(x)C(y)\,.
\end{split}
\end{equation}
As an immediate consequence of \eqref{eq:QB_commutator_master} the action of the Q-operator on a product of B-operators can be written as
\begin{equation}\label{eq:action_Q_on_Bstring}  
 \begin{split}
Q(x)B(x_{1})\cdots B(x_{m})& = q_m(x)B(x_{1})\cdots B(x_{m})W_{2m,0}(x) \\
 & \quad +\sum_{k=1}^{m} q_{k-1}(x)B(x_{1})\cdots B(x_{k-1})X_{2(k-1),0}(x,x_{k})B(x_{k+1})\cdots B(x_{m}) \,,
 \end{split}
\end{equation}   
where we defined the functions $q_k$ that are given by the polynomials
\begin{equation}\label{eq:Q_eigenvalue}
 q_k(x)=\prod_{i=1}^k(x-x_i)(x+x_i+1)\,.
\end{equation} 
They turn into Baxter q-functions when the parameters $x_i$ satisfy the Bethe equations $G_j^I=0$ as given in Section~\ref{sec:bae}. Further details of the computation are provided in the following subsections.

\subsection{Fundamental commutation relations for the Q-operator}\label{app:fcrQ} 

The following fundamental commutation relations are obtained from the boundary Yang-Baxter equation in \eqref{eq:BYBE-1}:
\begin{align}
 &   \dmonQ(x)A(y)+\bar{\mathbf{a}}\dmonQ(x)\mathbf{a}A(y)+\dmonQ(x)\bar{\mathbf{a}}C(y)+\bar{\mathbf{a}}\dmonQ(x)(x+y+1+\NN )C(y)\nonumber \\
 &=  A(y)\dmonQ(x)+A(y)\bar{\mathbf{a}}\dmonQ(x)\mathbf{a}+B(y)\mathbf{a}\dmonQ(x)+B(y)(x+y+1+\NN )\dmonQ(x)\mathbf{a}\,,\label{eq:BYBE_I}
\end{align}
and
\begin{align}
 &  \dmonQ(x)B(y)+\bar{\mathbf{a}}\dmonQ(x)\mathbf{a}B(y)+\dmonQ(x)\bar{\mathbf{a}}D(y)+\bar{\mathbf{a}}\dmonQ(x)(x+y+1+\NN )D(y)\nonumber \\
 & =  A(y)\dmonQ(x)\bar{\mathbf{a}}+A(y)\bar{\mathbf{a}}\dmonQ(x)(x-y+1+\NN ) +B(y)\mathbf{a}\dmonQ(x)\bar{\mathbf{a}}+B(y)(x+y+1+\NN )\dmonQ(x)(x-y+1+\NN )\,.\label{eq:BYBE_II}
\end{align} 
Similar as in the algebraic Bethe ansatz, we are looking for the  exchange relation of the creation operator $B(x)$ with the Q-operator, i.e. the diagonal terms of the double-row-monodromy $\dmonQ$, such that after the exchange, all operators $A$, $C$, and $D$ only appear on the right side of $\dmonQ$.  The operators $A$, $C$, and $D$  can then be commuted with the B-operators using the fundamental commutation relations of the transfer matrix listed in Appendix~\ref{sec:fcrT}.

\subsection{Exchange relation with the B-operator}
The exchange relation relevant to compute the action of the double-row  monodromy on B-operators and to diagonalise the Q-operator is 
\begin{equation}\label{eq:UB}
\tilde{\dmonQ}(x)B(y)  =  (x-y)(x+y+1)B(y)\sum_{n=0}^{\infty}(-1)^{n}\bar{\mathbf{a}}^{n}\dmonQ(x)\mathbf{a}^{n}+ F(x,y)\,,
\end{equation} 
where we introduced the abbreviations
\begin{equation}
 \tilde{\dmonQ}(x)  = \dmonQ(x)+\bar{\mathbf{a}}\dmonQ(x)\mathbf{a}\,,\label{eq:Mtilde_def}
\end{equation} 
that we call conjugated double-row-monodromy, and 
\begin{equation}
\begin{split}
  F(x,y)&=\left[ \tilde{\dmonQ}(x)\bar{\mathbf{a}}+(x-y)\bar{\mathbf{a}}\dmonQ(x)\right] A(y) -\left[ \tilde{\dmonQ}(x)\bar{\mathbf{a}}+(x+y)\bar{\mathbf{a}}\dmonQ(x)\right] D(y)\\
  &\quad\;+\left[ \tilde{\dmonQ}(x)\bar{\mathbf{a}}^{2}+2x\bar{\mathbf{a}}\dmonQ(x)\bar{\mathbf{a}}+(x-y)(x+y+1)\bar{\mathbf{a}}^{2}\sum_{n=0}^{\infty}(-1)^{n}\bar{\mathbf{a}}^{n}\dmonQ(x)\mathbf{a}^{n}\right] C(y)\,.
  \end{split}
\end{equation} 
We remark that in the expression above, all operators $A,B,C,D$ are to the right of the Q-operator double-row monodromy $\dmonQ$.
The derivation of the exchange relation \eqref{eq:UB} is shown in the remaining part of this subsection.

To derive \eqref{eq:UB} it is enough to consider the fundamental commutation relations \eqref{eq:BYBE_I} and \eqref{eq:BYBE_II}.
With the definition \eqref{eq:Mtilde_def} we rewrite \eqref{eq:BYBE_I} as
\begin{equation}
    \tilde{\dmonQ}(x)A(y)+\tilde{\dmonQ}(x)\bar{\mathbf{a}}C(y)+(x+y)\bar{\mathbf{a}}\dmonQ(x)C(y) =  A(y)\tilde{\dmonQ}(x)+B(y)\mathbf{a}\tilde{\dmonQ}(x)+(x+y)B(y)\dmonQ(x)\mathbf{a}\,,\label{eq:BYBE_I_mod}
\end{equation}
while (\ref{eq:BYBE_II}) becomes
\begin{align}
   \tilde{\dmonQ} (x)B(y)+\tilde{\dmonQ} (x)\bar{\mathbf{a}}D(y)&+(x+y)\bar{\mathbf{a}}\dmonQ (x)D(y)
  =  A(y)\tilde{\dmonQ} (x)\bar{\mathbf{a}}+(x-y)A(y)\bar{\mathbf{a}}\dmonQ (x)+B(y)\mathbf{a}\tilde{\dmonQ} (x)\bar{\mathbf{a}}\nonumber \\
 &  \quad\; +B(y)\mathbf{a}\bar{\mathbf{a}}\dmonQ (x)(x-y)+B(y)(x+y)\dmonQ (x)\mathbf{a}\bar{\mathbf{a}} +B(y)(x+y)\dmonQ (x)(x-y)\,.\label{eq:BYBE_II_mod}
\end{align}
Multiplying (\ref{eq:BYBE_I_mod}) by $\bar{\mathbf{a}}$ from the
right and subtracting it from (\ref{eq:BYBE_II_mod})   we arrive to the
relation
\begin{align} (x-y)(x+y+1)B(y)\dmonQ (x)+(x-y) \bar{\mathbf{a}} Z(x,y)&=
 \tilde{\dmonQ} (x)B(y)+\tilde{\dmonQ} (x)\bar{\mathbf{a}}D(y)+(x+y)\bar{\mathbf{a}}\dmonQ (x)D(y)\nonumber\\
 &\quad\;-\tilde{\dmonQ} (x)A(y)\bar{\mathbf{a}}-\tilde{\dmonQ} (x)\bar{\mathbf{a}}C(y)\bar{\mathbf{a}}-(x+y)\bar{\mathbf{a}}\dmonQ (x)C(y)\bar{\mathbf{a}}
  \,,\label{eq:BYBE_I_II_combined}
\end{align}
with
\begin{equation}\label{eq:Zett}
 Z(x,y)=A(y)\dmonQ (x)+B(y)\mathbf{a}\dmonQ (x)\,.
\end{equation} 
In \eqref{eq:BYBE_I_II_combined} the only term, where $A$ and $B$ needs to be commuted to the right side of $\dmonQ $ is denoted by $Z(x,y)$. This can be done as follows: First we note that the combination denoted by $Z(x,y)$ in \eqref{eq:Zett} of $A$ and $B$ appears in \eqref{eq:BYBE_I}. The latter equation can then  be reformulated as
\begin{align}
   \dmonQ (x)A(y)+\dmonQ (x)\bar{\mathbf{a}}C(y)-Z(x,y)
 & =  -\bar{\mathbf{a}}(\dmonQ (x)A(y)+\dmonQ (x)\bar{\mathbf{a}}C(y)-Z(x,y))\mathbf{a} \nonumber \\
 &  \quad\; +B(y)(x+y+1)\dmonQ (x)\mathbf{a}-\bar{\mathbf{a}}\dmonQ (x)(x+y+1)C(y)\,.
\end{align}
Solving this equation iteratively for $Z(x,y)$ we obtain
\begin{equation}\label{eq:BYBE_I_iter}
\begin{split}
 Z(x,y)&=   \dmonQ (x)A(y)+\dmonQ (x)\bar{\mathbf{a}}C(y)\\
 & \quad\;- \sum_{n=0}^{\infty}(-1)^{n}\bar{\mathbf{a}}^{n}\big(B(y)(x+y+1)\dmonQ (x)\mathbf{a}-\bar{\mathbf{a}}\dmonQ (x)(x+y+1)C(y)\big)\mathbf{a}^{n}\,.
 \end{split}
\end{equation}
This form of $Z(x,y)$ can then be inserted into \eqref{eq:BYBE_I_II_combined} and after some manipulation we arrive at \eqref{eq:UB} which we wanted to show.

\subsection{Q-operator and auxiliary operators}\label{sec:auxop}
The exchange relations \eqref{eq:QB_commutator_master} can now be derived  by multiplying  \eqref{eq:UB} with $ \frac{\Gamma(-p-x+i-\NN)}{\Gamma(-p-x+i+1)}\,\bar{\mathbf{a}}^{j}$   and subsequently taking the trace over the auxiliary space. A lengthy computation shows  that all trace operators on both sides of the resulting equation  can be expressed via $W_{i,j}$ in \eqref{eq:general_Q} and indeed reproduce \eqref{eq:QB_commutator_master}. 
This follows  after expressing all double-row monodromies $\dmonQ$ in terms of $\tilde{\dmonQ}(x)$ using the inverse relation
\begin{equation}\label{eq:M_Mtilde_inverse_relation}
\dmonQ(x) = \sum_{n=0}^{\infty}(-1)^{n}\bar{\mathbf{a}}^{n}\tilde{\dmonQ}(x)\mathbf{a}^{n}\,,
\end{equation}
that can be obtained from  \eqref{eq:Mtilde_def}. The resulting trace operators that contain infinite sums as the one appearing in \eqref{eq:M_Mtilde_inverse_relation}  are then rewritten using the identity 
\begin{equation} \label{eq:conjug}
\begin{split}
  \tr \frac{\Gamma(-p-x+i-\NN)}{\Gamma(-p-x+i+1)}\,\bar{\mathbf{a}}^{j}\sum_{n=0}^{\infty}(-1)^{n}\bar{\mathbf{a}}^{n}\mathcal{O}\mathbf{a}^{n}
  &=\tr \frac{\Gamma(- p-x+i+1-\NN)}{\Gamma(-p-x+i+2)}\,\bar{\mathbf{a}}^{j}\mathcal{O}\,,
\end{split}
\end{equation}
that holds for any operator $\mathcal{O}$.
The latter identity can be derived using the cyclicity of the trace, the relation 
\begin{equation}\label{eq:oscsum}
 \mathbf{a}^{n}\bar{\mathbf{a}}^{n}=\frac{\Gamma(\NN+n+1)}{\Gamma(\NN+1)}\,,
\end{equation} 
and the following identity for Gamma functions
\begin{equation}
\sum_{n=0}^{\infty}(-1)^{n}\frac{\Gamma(\alpha-n)\Gamma(\beta+n+1)}{\Gamma(\beta+1)}= \frac{\Gamma(1+\alpha)}{1+\alpha+\beta}\,,\label{eq:magic_sum_start}
\end{equation}
valid for arbitrary parameters $\alpha,\beta\in \mathbb{C}$.  In this way we obtain \eqref{eq:QB_commutator_master}.

Finally we note that combining \eqref{eq:M_Mtilde_inverse_relation} with the definition of the Q-operator \eqref{eq:qop} and the relation \eqref{eq:conjug} immediately leads to $Q(x)=W_{0,0}(x)$.

\section{Bethe ansatz for the Q-operator}\label{sec:Qonbethestate}

The action of the Q-operator on a Bethe off-shell state can be expressed by acting with the operatorial equation \eqref{eq:action_Q_on_Bstring} on the pseudo vacuum $| \Omega \rangle$. We immediately obtain
\begin{equation}
 \begin{split}
Q(x)B(x_{1})\cdots B(x_{m})|\Omega\rangle  & = q_m(x)B(x_{1})\dots B(x_{m})W_{2m,0}(x)|\Omega\rangle  \\
 & \quad +\sum_{k=1}^{m} q_{k-1}(x)B(x_{1})\dots B(x_{k-1})X_{2(k-1),0}(x,x_{k})B(x_{k+1})\dots B(x_{m})|\Omega\rangle   \,.\label{eq:action_Q_on_Bstate}  
 \end{split}
\end{equation} 
The first line on the right hand side is the wanted term that comprises the  eigenvalues and eigenvectors of the Q-operator. As we can see, we obtain the expected eigenvalue $q_m(x)$ given in \eqref{eq:Q_eigenvalue} and an additional normalisation 
arising from the action of $W_{2m,0}$ on the reference state:
\begin{equation}\label{eq:normal}
 W_{2m,0}(x)|\Omega\rangle =
\frac{ 1 }{2m- p-q-2Ns}|\Omega\rangle\,,
\end{equation}  
 cf.~\eqref{eq:result}. The equation \eqref{eq:normal} above we obtain in Section~\ref{sec:Wanted_terms}. The second line of the right hand side of \eqref{eq:action_Q_on_Bstate} contains the so called unwanted terms. In Section~\ref{sec:unwanted} we compute the relevant part of the coefficients in front of the unwanted terms, and in Section~\ref{sec:vanishing} we are going to show that they vanish if  the Bethe ansatz equations $G_j^I=0$ for all $j\in I$ are satisfied. The function $G_j^I$ was defined in \eqref{eq:GG}.

\subsection{Wanted term}\label{sec:Wanted_terms}

In this section we compute the action of $W_{2m,0}$ on the reference state which is given by the $N$-fold tensor product of highest weight states 
\begin{equation}
 |\Omega\rangle=|\text{hws}\rangle\otimes\ldots \otimes |\text{hws}\rangle\,,
\end{equation} 
characterised by the properties that at each site $S_+|\text{hws}\rangle=0$ and $S_3|\text{hws}\rangle=s|\text{hws}\rangle$.
For this purpose it is convenient to write the operator $W_{2m,0}$  in terms of the double-row monodromy using relation \eqref{eq:conjug}, such that
\begin{equation}
W_{2m,0}(x) 
=\tr \frac{\Gamma(-p-x+2m-1-\NN)}{\Gamma(-p-x+2m)}  \dmonQ(x)
\end{equation} 
From this rewriting it follows that the operator $W_{2m,0}$ is related to the Q-operator \eqref{eq:qop} by a shift of $2m$ in  the parameter $p$ which only appears in the boundary K-matrix.
As the Q-operator commutes by construction with the transfer matrix, it is like the transfer matrix block diagonal and does not mix states with different magnon numbers.  It follows that also $W_{2m,0}$ is block diagonal and thus the action of $W_{2m,0}$ on the reference state is diagonal. 

We now turn to the computation of the eigenvalue. First we note that the action of the Lax operators in \eqref{rop1} that appear in the monodromy of the Q-operator  on the local highest-weight states can simply written as
\begin{equation}
 \mathcal{R}_+(x)|\text{hws}\rangle =e^{\oa S_- }|\text{hws}\rangle \,,\qquad \langle\text{hws}|\mathcal{R}_+(x)=\langle \text{hws}|e^{\oad S_+}\,.
\end{equation}  
It thus follows that
 \begin{equation}
\begin{split}
\langle \Omega| W_{2m,0}(x)|\Omega\rangle& =\tr \frac{\Gamma(-p-x+2m-1-\NN)}{\Gamma(-p-x+2m)}  \langle \Omega| e^{\oad S_+^{[1]}}\cdots e^{\oad S_+^{[N]}} \KQR_+(x)e^{\oa S_-^{[N]}}\cdots e^{\oa S_-^{[1]}} |\Omega\rangle \,.
\end{split}
\end{equation} 
When expanding the exponents that arise from  the two single-row monodromies,  we see that at each site only the terms with equal powers contribute. 
Using the relation
\begin{equation}
\begin{split}
\langle \Omega|  \prod_{j=1}^N\left(S_+^{[j]}\right)^{k_j}\left(S_-^{[j]}\right)^{k_j}|\Omega\rangle= \prod_{j=1}^N\frac{(2s)!k_j!}{(2s-k_j)!}
\end{split}
\end{equation} 
we then obtain
\begin{equation}
\begin{split}
\langle \Omega| W_{2m,0}(x)|\Omega\rangle& =\sum_{k_1,\ldots,k_N=0}^\infty    \prod_{j=1}^N\binom{2s}{k_j}\tr\frac{\Gamma(-p-x+2m-1-\NN)}{\Gamma(-p-x+2m)} \oad^{k_1+\ldots+k_N} \KQR_+(x)  \oa^{k_1+\ldots+k_N} \,.
\end{split}
\end{equation} 
Using the cyclicity of the trace and formula \eqref{eq:oscsum} we further simplify this expression and arrive at 
\begin{equation}
\begin{split}
\langle \Omega| W_{2m,0}(x)|\Omega\rangle& =\sum_{k_1,\ldots,k_N=0}^\infty    \prod_{j=1}^N\binom{2s}{k_j}\tr \frac{\Gamma(- p-z+2m-\NN-1)\Gamma( q-z)}{\Gamma(-p-z+2m)\Gamma( q-z-\NN+k_{tot})} \frac{\Gamma(\NN+1)}{\Gamma(\NN-k_{tot}+1)}   \,.
\end{split}
\end{equation} 
where $k_{tot}=\sum_{i=1}^Nk_i$. The trace can the be taken using the relation
\begin{equation}
\begin{split}
 \sum_{n=k_{tot}}^\infty& \frac{\Gamma(- p-x+2m-n-1)\Gamma( q-x)}{\Gamma(-p-x+2m)\Gamma( q-x-n+k_{tot})} \frac{\Gamma(n+1)}{\Gamma(n-k_{tot}+1)}   =-(-1)^{k_{tot}}k_{tot}!
\frac{ \Gamma (p+q-2 m) }{ \Gamma (p+q-2m+k_{tot}+1)}
 \end{split}
\end{equation} 
This can be shown by rewriting the sum arising from the trace into a  function proportional to the  hypergeometric function $_2F_1$ at $x=1$  which can then be  expressed in terms of Gamma functions. Remarkably the $x$ dependence cancels out and we find
\begin{equation}
\begin{split}
\langle \Omega| W_{2m,0}(x)|\Omega\rangle& = -\sum_{k_1,\ldots,k_N=0}^\infty    \prod_{j=1}^N\binom{2s}{k_j}(-1)^{k_{tot}}k_{tot}!
\frac{ \Gamma (p+q-2 m) }{ \Gamma (p+q-2m+k_{tot}+1)}\,.
\end{split}
\end{equation}  
The remaining sums can then be evaluated successively using  the relation
\begin{equation}
 \sum_{k=0}^\infty  \binom{2s}{k}(-1)^{k}(k+w)!
\frac{ \Gamma (p+q-2 m) }{ \Gamma (p+q-2m+k+w+1)}=\frac{\Gamma (w+1) \Gamma (p+q-2 m+2 s)}{\Gamma (p+q-2 m+2 s+w+1)}\,.
\end{equation} 
we can evaluate all sums successively and we recover \eqref{eq:normal}.

\subsection{Unwanted terms}\label{sec:unwanted}

In this section we bring the unwanted terms in \eqref{eq:action_Q_on_Bstate} to a form that is convenient to show that they indeed vanish, as shown in Section~\ref{sec:vanishing}. 
The logic is as follows. We take the definition of $X_{i,j}$ \eqref{eq:X_definition}, and commute all operators contained in it, i.e. $A, \tilde{D}$, and $C$  together with the auxiliary operators $W_{i,j}$, to the right of the $B$ operators in \eqref{eq:action_Q_on_Bstate}. The exchange relations are found in  \eqref{eq:commute_AB}, \eqref{eq:commute_DB}, \eqref{eq:commute_CB2}, and \eqref{eq:QB_commutator_master}. The commutation relations involving $W_{i,j}$ reintroduce an operator $X$, see \eqref{eq:QB_commutator_master}, and hence we need to iterate the procedure above.
In the final step, we use that the operators  $A$ and $\tilde D$ act diagonally on the pseudovacuum, see \eqref{eq:actref}, while $C$ annihilates it. 
The subspace of unwanted terms is then spanned by states that are given by a string of $B$ operators and an auxiliary operator $W$ acting on the pseudovacuum.  

The $B$ operators commute among themselves and consequently  the off-shell Bethe states are symmetric under the permutations of the Bethe roots.
Due to this symmetry it is sufficient  to focus on the coefficients of states, which do not involve the action of the operator $B(x_1)$  to conclude  the vanishing of all the unwanted terms. 
All the unwanted terms without
$B(x_{1})$ are contained in the term with summation index $k=1$ of \eqref{eq:action_Q_on_Bstate}, i.e. the contribution
\begin{equation}
X_{0,0}(x,x_{1})|I_1\rangle \qquad \text{with}\qquad 
    I_1=(2,\ldots,m)\,,
\end{equation} 
the ordered set without $1$. 
In the following we show that all $B(x_1)$-independent terms can be written as
\begin{equation}\label{eq:Xact}
X_{0,0}(x,x_{1}) |I_1\rangle\simeq \sum_{n=0}^{|I_1|}\sum_{\substack{J' \subseteq I_1\\
|{J'}|=n}}\mathcal{G}^{n}(0,I_1,{J'})W_{n+1,n+1}(x)\left|I_1\setminus {J'}\right\rangle \,.
\end{equation}
The symbol  $\simeq$ indicates the omission of terms with $B(x_1)$ dependence.  The coefficients $\mathcal{G}^{n} $ satisfy a recursion relation that we will now derive and solve to find their explicit form, see \eqref{eq:G_sol}.

To derive \eqref{eq:Xact} let us consider the action $X_{i,j}(x,x_{1})\left|J\right\rangle$ of the operator $X$ on an  ordered subset $J\subseteq I_1$. 
When commuting the $A(x_1)$, $\tilde{D}(x_1)$, and $C(x_1)$ operators to the right of the string of $B$ operators, the commutation relations \eqref{eq:commute_AB}, \eqref{eq:commute_DB} and \eqref{eq:commute_CB2}  can reintroduce $B(x_1)$ dependence for some terms. Whenever this happens we commute the $W_{i',j'}(x)$ from $X_{i,j}(x,x_{1})$ over $B(x_1)$ and subsequently omit the remaining  $B(x_1)$-dependent terms. In this way we arrive at the recursion relation 
\begin{equation}\label{eq:X_recursion}
 \begin{split}
X_{i,j}(x,x_{1})\left|J\right\rangle  & \simeq  \left( (-p-x_1+i){\mathcal{A}}_{1}^{J}+(p-x_1-1-i)\tilde{\mathcal{D}}_{1}^{J}\right) W_{i+1,j+1}(x)\left|J\right\rangle \\
 &\quad\;   -(-p-x_1+i+1)(p-x_1-2-i)\sum_{k\in J}\mathcal{C}_{1,k}^{J }W_{i+2,j+2}(x)\left|J\setminus  k  \right\rangle \\
 &  \quad\; +\sum_{k\in J}\left( (-p-x_1+i){\mathcal{A}}_{1,k}^{J }+(p-x_1-1-i)\tilde{\mathcal{D}}_{1,k}^{J }\right) X_{i+1,j+1}(x,x_{1})\left|J\setminus k \right\rangle \\
 & \quad\;  -\sum_{\substack{ k,l\in  J\\k<l}}(-p-x_1+i+1)(p-x_1-2-i)\mathcal{C}_{1,k,l}^{J }X_{i+2,j+2}(x,x_{1})\left|J\setminus\left ( k,l\right) \right\rangle \,,
 \end{split}
\end{equation}  
where the $\mathcal{A}$, $\tilde{\mathcal{D}}$, and $\mathcal{C}$ coefficients are defined in Appendix~\ref{app:comm}.

From the recursion \eqref{eq:X_recursion} we see that the action of the operator $X_{i,j}$ on a given state $\left|J\right\rangle $ can be expanded as
\begin{equation} \label{eq:X_action_expansion1}
X_{i,j}(x,x_{1})\left|J\right\rangle  \simeq \sum_{n=0}^{|J|}\sum_{\substack{J' \subseteq J\\
|{J'}|=n}}\mathcal{G}^{n}(i,J,{J'})W_{n+i+1,n+j+1}(x)\left|J\setminus {J'}\right\rangle \,,
\end{equation}
where $|J|$ and $|{J'}|$ are the cardinalities of the ordered sets
$J$ and ${J'}$, and the $x_{1}$ dependence of the coefficients
$\mathcal{G}^{n}$ is suppressed in our notation. Substituting the
expansion \eqref{eq:X_action_expansion1} into \eqref{eq:X_recursion}
gives a recursion relation for the coefficients. It reads
\begin{equation}\label{eq:G_recursion}
 \begin{split} 
\mathcal{G}^{n}(i,J,{J'}) & =  \sum_{k\in {J'}}\left((-p-x_1+i){\mathcal{A}}_{1,k}^{J }+(p-x_1-1-i)\tilde{\mathcal{D}}_{1,k}^{J }\right) \mathcal{G}^{n-1}(i+1,J\setminus k ,{J'}\setminus k )\\
 &   -\sum_{\substack{k,l \in  J'\\k<l}}(-p-x_1+i+1)(p-x_1-2-i)\mathcal{C}_{1,k,l}^{J }\,\mathcal{G}^{n-2}(i+2,J\setminus(k,l) ,{J'}\setminus\left( k,l\right) )\,,
 \end{split}
\end{equation} 
for $n\ge2$, ${J'}\subset J$ and $|{J'}|=n$, with the initial
conditions
\begin{equation}\label{eq:G_initial_1}
\mathcal{G}^{0}(i,J,\emptyset) =  (-p-x_1+i){\mathcal{A}}_{1}^{J}+(p-x_1-1-i)\tilde{\mathcal{D}}_{1}^{J}\,,
\end{equation} 
for $n=0$ and
\begin{equation}\label{eq:G_initial_2}
\begin{split}
 \mathcal{G}^{1}(i,J,(k) ) & =  \left(  (-p-x_1+i){\mathcal{A}}_{1,k}^{J}+(p-x_1-1-i)\tilde{\mathcal{D}}_{1,k}^{J }\right) \mathcal{G}^{0}(i+1,J\setminus  k ,\emptyset)\\
 &\qquad -(-p-x_1+1+i)(p-x_1-2-i)\mathcal{C}_{1,k}^{J }\,,
 \end{split}
\end{equation}
with $k\in I_1$ and $n=1$.
The full solution of the recursion (\ref{eq:G_recursion}) is 
\begin{equation} \label{eq:G_sol}
\begin{split}
\mathcal{G}^{n}(i,J ,{J'})  &=  \frac{\Gamma(-p-x_{1}+1+n+i)}{\Gamma(-p-x_{1})}{\mathcal{A}}_{1}^{J }\sum_{\sigma\in S_{n}}\prod_{k=1}^{n}{\mathcal{F}}_{\mathcal{A}}\left(J ,J'_{\sigma,k}\right)\\
&\quad
+\frac{\Gamma(p-x_{1})}{\Gamma(p-x_{1}-n-1-i)}\tilde{\mathcal{D}}_{1}^{J}\sum_{\sigma\in S_{n}}\prod_{k=1}^{n}{\mathcal{F}}_{\tilde{ \mathcal{D}}}\left(J ,J'_{\sigma,k} \right)\,,
\end{split}
\end{equation}  
with $n\in\mathbb{N}$ and
\begin{equation} \label{eq:FA_coeffs}
 {\mathcal{F}}_{\mathcal{A}}(J ,J'_{\sigma,k} ) =  \frac{{\mathcal{A}}_{j'_{\sigma(k)}}^{J\setminus  J'_{\sigma,k}} }{x_{1}-x_{j'_{\sigma(k)}} -1}-\frac{\tilde{\mathcal{D}}_{j'_{\sigma(k)}}^{J\setminus  J'_{\sigma,k}} }{x_{1}+x_{ j'_{\sigma(k)}} }\,,
\end{equation} 
\begin{equation}\label{eq:FD_coeffs}
 {\mathcal{F}}_{\tilde{\mathcal{D}}}(J ,J'_{\sigma,k} ) =  \frac{{\mathcal{A}}_{j'_{\sigma(k)}}^{J\setminus  J'_{\sigma,k}} }{x_{1}-x_{j'_{\sigma(k)}} +2}-\frac{\tilde{\mathcal{D}}_{j'_{\sigma(k)}}^{J\setminus  J'_{\sigma,k}} }{x_{1}-x_{ j'_{\sigma(k)}} +1}\,.
\end{equation}   
The sums in \eqref{eq:G_sol} go over all permutations $S_n$. Further we introduced  the notation
\begin{equation}
  J'_{\sigma,k}=\left({j'}_{\sigma(1)},{j'}_{\sigma(2)},\dots,{j'}_{\sigma(k)}\right)\,.
\end{equation} 
It can easily be seen that  \eqref{eq:G_sol}  yields \eqref{eq:G_initial_1} and \eqref{eq:G_initial_2} for $n=0,1$. Showing that  \eqref{eq:G_sol}  satisfies the  general recursion relation \eqref{eq:G_recursion} for $n>1$ can be done inductively but is more involved.

\subsection{Vanishing of the unwanted terms}\label{sec:vanishing}

In this section we show the coefficients \eqref{eq:G_sol} vanish in case the Bethe ansatz equations \eqref{eq:GG} are satisfied. It is easy to see for the coefficients with $n=0$. The coefficient $\mathcal{G}^{0}(0,I_1,\emptyset)$ is equal to  $G_1^I$ in the Bethe equations defined in \eqref{eq:GG}, hence it trivially vanishes when the Bethe equations are satisfied. Its permutations in the Bethe roots give the remaining $G_k^I$ and therefore also vanish. The ordered sets in the formulae are related as $I=I_1\cup 1=\{1,\ldots,m\}$. 

Showing that the coefficients $\mathcal{G}$ vanish for $n>0$ is more involved.
 For this purpose we note that they can be rewritten as
\begin{equation}\label{eq:altform}
 \begin{split}
  \mathcal{G}^n(0,I_1,J')&= (-1)^n\left(\mathcal{X}_n({x_1})-\mathcal{X}_n(-x_1-1)\right)  \frac{\Gamma(-p-x_1+1+n)}{\Gamma(-p-x_1)}  \frac{\Gamma(p-x_1-1)}{\Gamma(p-x_1-n-1)}\mathcal{A}^I_1 \\
  &\qquad\quad\times\prod_{\substack{l,k\in J'\cup 1\\ l<k }}\frac{1}{{\bf f}(x_k,x_l){\bf h}(x_k,x_l)}
  \prod_{k\in J'}\frac{(1+2x_k) \mathcal{A}^I_k }{(x_k-x_1)(x_k+x_1+1)(p-x_k-1)}\,,
 \end{split}
\end{equation} 
with
\begin{equation}\label{eq:x}
\begin{split}
 \mathcal{X}_{n}  (x_1)=  \frac{\Gamma\left(p-x_{1}-n-1\right)}{\Gamma\left(p-x_{1}-1\right)}\sum_{\sigma\in S_{n}}\prod_{k=1}^{n}\frac{1}{1+2x_{{j'}_{\sigma\left(k\right)}}}&\left[(p-x_{{j'}_{\sigma\left(k\right)}}-1)(x_{{j'}_{\sigma\left(k\right)}}+x_{1}+2)\prod_{l\in J'_{\sigma,k-1} }\mathbf{h}\left(x_{{j'}_{\sigma\left(k\right)}},x_{l}\right)\right.\\
 & \; \left.+\,(p+x_{{j'}_{\sigma\left(k\right)}})(x_{{j'}_{\sigma\left(k\right)}}-x_{1}-1)\prod_{l\in J'_{\sigma,k-1} }\mathbf{f}\left(x_{{j'}_{\sigma\left(k\right)}},x_{l}\right)\right].
 \end{split}
\end{equation}  
To derive this expression, we take the solution \eqref{eq:G_sol} with $J=I_1$, and  relate the ${\mathcal{A}}_{j'_{\sigma(k)}}^{I_1\setminus  J'_{\sigma,k}}$ and ${\tilde{\mathcal{D}}}_{j'_{\sigma(k)}}^{I_1\setminus  J'_{\sigma,k}}$ terms from \eqref{eq:FA_coeffs} and \eqref{eq:FD_coeffs} to ${\mathcal{A}}_{j'_{\sigma(k)}}^{I}$ and ${\tilde{\mathcal{D}}}_{j'_{\sigma(k)}}^{I}$ through combinations of the functions  $\mathbf{f}(x_i,x_j)$ and $\mathbf{h}(x_i,x_j)$ in \eqref{eq:letters}, cf.~\eqref{eq:hfhf}. After substituting all ${\tilde{\mathcal{D}}}_{j'_{\sigma(k)}}^{I}$ with ${{\mathcal{A}}}_{j'_{\sigma(k)}}^{I}$  using the Bethe Ansatz equations $G_{j}^I=0$, see  \eqref{eq:GG}, we arrive at \eqref{eq:altform}.

Still it remains to show that \eqref{eq:altform} actually vanishes. This relies on some remarkable simplification of \eqref{eq:x}. More precisely we find that the finite sum is independent of $x_1$ and also all other Bethe roots and simply yields
 \begin{equation}
   \mathcal{X}_{n}(x_1)=n!\,.
 \end{equation} 
This can be easily checked for small  numbers $n$ and proved by induction in general.   Due to the tediousness of the calculation we do not provide all the details, but only the necessary ingredients.
The first step is to notice the  recursive relation
\begin{equation}
\begin{split}
 \mathcal{X}_{n} (x_1) =   \mathcal{X}_{n-1}(x_1)\sum_{k=1}^{n}\frac{1}{\left(1+2x_{{j'}_{k}}\right)\left(p-x_{1}-n-1\right)}&\left[(p-x_{{j'}_{k}}-1)(x_{{j'}_{k}}+x_{1}+2)\prod_{l\in J'\setminus{(k)} }\mathbf{h}\left(x_{{j'}_{k}},x_{l}\right)\right.\\
 & \quad \left.+\,(p+x_{{j'}_{k}})(x_{{j'}_{k}}-x_{1}-1)\prod_{l\in J'\setminus{(k)} }\mathbf{f}\left(x_{{j'}_{k}},x_{l}\right)\right]\,.
 \end{split}
\end{equation}  
This form invites for a proof  by  induction, and we only need to show that the sum over $k$ evaluates to $n$.
After relabeling the Bethe roots the equation that has to be  shown is
\begin{equation}\label{eq:magaum}
\begin{split}
\sum_{i=1}^{n}\left[\frac{(p+x_{i})(x_{i}-x_{0}-1)}{(1+2x_{i}) (p-x_{0}-n-1) }\prod_{j\neq i}^{n}\mathbf{f}(x_{i},x_{j})+\frac{(p-x_{i}-1)(x_{i}+x_{0}+2)}{(1+2x_{i} )   (p-x_{0}-n-1)}\prod_{j\neq i}^{n}\mathbf{h}(x_{i},x_{j})\right] & =  n\,.
\end{split}
\end{equation}
This we  prove by another induction. The calculation is straightforward and depends on the partial fraction decompositions 
\begin{align}
\prod_{i=1}^{n}\mathbf{f}(x,x_{i}) & =  1-\sum_{i=1}^{n}\frac{2x_{i}}{1+2x_{i}}\frac{\prod_{j\neq i}^{n}\mathbf{f}(x_{i},x_{j})}{x-x_{i}}-\sum_{i=1}^{n}\frac{2(1+x_{i})}{1+2x_{i}}\frac{\prod_{j\neq i}^{n}\mathbf{h}(x_{i},x_{j})}{x+x_{i}+1}\,,\\
\prod_{i=1}^{n}\mathbf{h}(x,x_{i}) & =  1+\sum_{i=1}^{n}\frac{2x_{i}}{1+2x_{i}}\frac{\prod_{j\neq i}^{n}\mathbf{f}(x_{i},x_{j})}{x+x_{i}+1}+\sum_{i=1}^{n}\frac{2(1+x_{i})}{1+2x_{i}}\frac{\prod_{j\neq i}^{n}\mathbf{h}(x_{i},x_{j})}{x-x_{i}}\,,
\end{align}
and the identity
\begin{equation}\label{eq:secsum}
\begin{split}
\sum_{i=1}^{n}\left[\prod_{j\neq i}^{n}\mathbf{f}(x_{i},x_{j})\frac{x_{i}}{(1+2x_{i})}+\prod_{j\neq i}^{n}\mathbf{h}(x_{i},x_{j})\frac{(1+x_{i})}{(1+2x_{i})}\right] & =  n\,.
\end{split}
\end{equation}
The partial fraction decompositions can again be shown by the steps of induction using the form of the functions $\mathbf{f}$ and $\mathbf{h}$ from \eqref{eq:letters}. The sum \eqref{eq:secsum} is also easily proven by induction and the help of the partial fraction decomposition relations. The sums in \eqref{eq:magaum} and \eqref{eq:secsum}  may be seen as the generalisation of the magic sum in \cite{Frassek:2015qra} equation (C.8) for the closed chain.

Finally, since $\mathcal{X}_{n}(x_1)=\mathcal{X}_{n}(-x_1-1)$ all  $\mathcal{G}^n(0,I_1,J')$ coefficient vanish as a consequence of the Bethe ansatz equations. Thus we conclude that due to the symmetry in the Bethe roots, all the unwanted term vanish.

\section{Conclusions}\label{sec:conc} 
In this paper we constructed the Q-operators for the open spin $s$ XXX Heisenberg chain with diagonal boundaries and diagonalised them using the algebraic Bethe ansatz. We showed the convergence of the trace in the infinite-dimensional auxiliary space by computing the eigenvalues arising from the wanted term. Furthermore we showed explicitly that the unwanted terms vanish if the Bethe equations are satisfied. 

We expect that our results can be generalised to the case of the open XXZ spin chain for which the Q-operators were introduced in \cite{Baseilhac:2017hoz}.   
It would further be interesting to see whether the Bethe ansatz for Q-operators studied in this paper can provide us with new determinant formulas for scalar products by following the ideas of \cite{Belliard:2019bfz} that rely on the explicit form of the unwanted terms. 
Finally, the Bethe ansatz for Q-operators as done here is quite involved and one may suspect that there is an alternative way of doing the Bethe ansatz where the B-operators of the transfer matrix monodromy are substituted by their analogs arising from the Q-operator monodromy. In the case of higher spin transfer matrices we refer the reader to \cite{tarasov1988algebraic,Melo:2008okn,Martins:2009dt,Lima-Santos:2017bpy}.

\section*{Acknowledgements}
We thank Samuel Belliard, Mikhail Isachenkov, Rodrigo A. Pimenta and Robert Weston  for interesting discussions. IMSZ thanks the University of Modena and Reggio Emilia for warm hospitality during the final stage of finishing this project. 
The research leading to these results has received funding from the People Programme
(Marie Curie Actions) of the European Union’s Seventh Framework Programme FP7/2007-
2013/ under REA Grant Agreement No 317089 (GATIS). 
RF received support of the German research foundation (DFG)
Research Fellowships Programme 416527151.
The work of IMSZ was supported by the grant “Exact Results in Gauge and String Theories” from the Knut and Alice Wallenberg foundation. IMSZ  received support from Nordita that is supported in part by NordForsk.

\appendix

\section{Fundamental Commutation relation for the transfer matrix}\label{sec:fcrT}

In this appendix we provide the known fundamental commutation relations and eigenvalues on the reference state, see e.g. \cite{Belliard:2013aaa}.
The fundamental commutation relations arising from the boundary Yang-Baxter equation \eqref{eq:BYBE-A} needed for the algebraic Bethe ansatz discussed in Section~\ref{sec:bae} are 
\begin{equation} \label{eq:commute_AB}
A(x)B(y) =  \mathbf{f}(x,y)B(y)A(x)+\mathbf{g}_{A}(x,y)B(x)A(y)+\mathbf{g}_{\tilde{D}}(x,y)B(x)\tilde{D}(y)\,,
\end{equation}
and
\begin{equation}\label{eq:commute_DB}
\tilde{D}(x)B(y) = \mathbf{h}(x,y)B(y)\tilde{D}(x)+\mathbf{k}_{A}(x,y)B(x)A(y)+\mathbf{k}_{\tilde{D}}(x,y)B(x)\tilde{D}(y)\,.
\end{equation}
Here we defined the functions
\begin{equation}\label{eq:letters}
{ \displaystyle
\begin{array}{lll}
\mathbf{f}(x,y)=\frac{(x+y)(x-y-1)}{(x-y)(x+y+1)}\,, & \mathbf{g}_{A}(x,y)=\frac{2y}{(x-y)(1+2y)}\,, & \mathbf{g}_{\tilde{D}}(x,y)=\frac{-1}{(x+y+1)}\,,\\
\mathbf{h}(x,y)=\frac{(x-y+1)(x+y+2)}{(x-y)(x+y+1)}\,, & \mathbf{k}_{A}(x,y)=\frac{4y(1+x)}{(1+2x)(1+2y)(x+y+1)}\,, & \mathbf{k}_{\tilde{D}}(x,y)=\frac{-2(1+x)}{(x-y)(1+2x)}\,.
\end{array} }
\end{equation}
It then follows that the action on a Bethe off-shell state can be written as
\begin{equation}\label{eq:acta}
 A(x)\left|I\right\rangle    =  \alpha(x)\prod_{k=1}^m\mathbf{f}(x,x_{k})|I\rangle +\sum_{j\in I}\left[\frac{{\mathcal{A}}^{I}_j}{x-x_{j}}-\frac{\tilde { \mathcal{D}}^{I}_j}{1+x+x_{j}}\right]B(x)\left|I\setminus\left\{ j\right\} \right\rangle\,,
\end{equation} 
and
\begin{equation}\label{eq:actd}
 \tilde{D}(x)\left|I\right\rangle    =  {\tilde\delta}(x)\prod_{k=1}^m\mathbf{h}(x,x_{k}) |I\rangle + \frac{2+2x}{1+2x}\sum_{j\in I}\left[\frac{{\mathcal{A}}^{I  }_j}{1+x+x_{j}}-\frac{\tilde {\mathcal{D}}^{I}_j}{x-x_{j}}\right]B(x)\left|I\setminus\left\{ j\right\}  \right\rangle \,,
\end{equation}  
with the functions $\mathcal{A}^I_j$ and $\tilde{\mathcal{D}}^I_j$ defined in \eqref{eq:adcof}. We further not that the commutation relations between $B$ and $C$ are of the form
\begin{equation}\label{eq:commute_CB1}
\begin{split}
[C(x),B(y)] & =\frac{1}{1+x+y}\left(A(y)A(x)-D(x)D(y)\right)-\frac{x+y}{(x-y)(x+y+1)}\left(A(x)D(y)-A(y)D(x)\right)\,.
 \end{split}
\end{equation}
This relation can be written in terms of $\tilde D$ as
\begin{equation}\label{eq:commute_CB2}
\begin{split}
[C(x),B(y)]  &= \mathbf{l}_{\tilde{D}\tilde{D}}(x,y)\tilde{D}(x)\tilde{D}(y)+\mathbf{l}_{AA}(x,y)A(x)A(y)+\mathbf{m}_{AA}(x,y)A(y)A(x)\\
 &  \qquad +\mathbf{m}_{A\tilde{D}}(x,y)A(y)\tilde{D}(x)+\mathbf{l}_{A\tilde{D}}(x,y)A(x)\tilde{D}(y)+\mathbf{l}_{\tilde{D}A}(x,y)\tilde{D}(x)A(y)\,,
 \end{split}
\end{equation}
with the functions
\begin{equation}
{ \displaystyle
\begin{array}{lll} 
\mathbf{l}_{\tilde{D}\tilde{D}}(x,y)=\frac{-1}{1+x+y}\,, & \mathbf{m}_{AA}(x,y)=\frac{2x(x-y+1)}{(x-y)(1+2x)(1+x+y)}\,, & \mathbf{l}_{AA}(x,y)=\frac{-2x}{(x-y)(1+2x)(1+2y)}\,,\\
\mathbf{l}_{A\tilde{D}}(x,y)=\frac{-2x}{(x-y)(1+2x)}\,, & \mathbf{m}_{A\tilde{D}}(x,y)=\frac{(x+y)}{(x-y)(x+y+1)}\,, & \mathbf{l}_{\tilde{D}A}(x,y)=\frac{-1}{(1+2y)(x+y+1)}\,.
\end{array} }
\end{equation} 
For completeness we also give the action on the reference state of the operators $A$ and $\tilde D$ in \eqref{eq:actref}. The eigenvalues read
\begin{equation}\label{eq:ADvac}
    \alpha(x)=(q+x)\left(x+\frac{1}{2}+s\right)^{2N} \,,\qquad      \tilde \delta(x)=\frac{2x}{2x+1}(q-x-1)\left(x+\frac{1}{2}-s\right)^{2N}\,.
\end{equation}

\section{Action of $A$, $C$, and $\tilde{D}$ on Bethe states }\label{app:comm}

The action of $A$ and $\tilde{D}$ on a given state $\left|J\right\rangle$ with the ordered set $J \subseteq \left( 2,3,\dots,m\right)$ is given by
\begin{align} 
\frac{2x_{1}}{1+2x_{1}}A(x_{1})\left|J\right\rangle  & = {\mathcal{A}}_{1}^{J}\left|J\right\rangle +\sum_{j\in J}\tilde{\mathcal{A}}_{2}^{J\setminus j }(x_{1}|x_{j})\left|J\setminus j \cup 1 \right\rangle \,,\\
\tilde{D}(x_{1})\left|J\right\rangle  & =  \tilde{\mathcal{D}}_{1}^{J}(x_{1})\left|J\right\rangle +\sum_{j\in J}\tilde{\mathcal{D}}_{2}^{J\setminus j }(x_{1}|x_{j})\left|J\setminus j \cup 1 \right\rangle \,,
\end{align}
where the coefficients are
\begin{equation}\label{eq:hfhf}
 {\mathcal{A}}_{k}^{J}=  \frac{2x_k\alpha(x_k)}{1+2x_k}\prod_{j\in J\setminus k}\mathbf{f}(x_k,x_{j})\,,\qquad 
\tilde{\mathcal{D}}_{k}^{J}(x) =  \tilde{\delta}(x)\prod_{j\in J\setminus k}\mathbf{h}(x_k,x_{j})\,.
\end{equation} 
and
\begin{align}
{\mathcal{A}}_{k,l}^{J } & =  \frac{2x_{k}}{1+2x_{k}}\left[\frac{{\mathcal{A}}_l^{J\setminus k }}{x_{k}-x_{l}}-\frac{\tilde{\mathcal{D}}_l^{J\setminus k }}{1+x_{k}+x_{l}}\right] \,,\label{eq:Coeff_alpha_2}\\
\tilde{\mathcal{D}}_{k,l}^{J }  & =  \frac{2+2x_{k}}{1+2x_{k}}\left[\frac{{\mathcal{A}}_l^{J\setminus  k  } }{1+x_{k}+x_{l}}-\frac{\tilde{\mathcal{D}}_l^{J\setminus  k  }}{x_{k}-x_{l}}\right] \,,\label{eq:Coeff_delta_2}
\end{align} 
The action of the operator $C$ is given through
\begin{equation}
 C(x_{1})\left|J\right\rangle  = \sum_{j\in J}\mathcal{C}_{1,j}^{J }\left|J\setminus  j  \right\rangle +\sum_{\substack{j,k\in J\\j<k}}\mathcal{C}_{1,j,k}^{J }\left|J\setminus\left(j,k\right) \cup 1 \right\rangle \,,
\end{equation} 
with 
\begin{equation}
\mathcal{C}_{1,j}^{J } =\frac{1+2x_{1}}{2+2x_{1}}{\mathcal{A}}_{1}^{J\setminus  j  }\tilde{\mathcal{D}}_{1,j}^{J }+\frac{1+2x_{1}}{2x_{1}}\tilde{\mathcal{D}}_{1}^{J\setminus  j  }{\mathcal{A}}_{1,j}^{J } \,,\label{eq:Coeff_c_1}
\end{equation}
and
\begin{equation}\label{eq:Coeff_c_2_T}
\mathcal{C}_{1,j,k}^{J}  =\frac{1+2x_{1}}{2+2x_{1}} \left[ \mathcal{A}_{1,j}^{J\setminus k }-\frac{1}{2+2x_{1}}\tilde{\mathcal{D}}_{1,j}^{J\setminus k }\right]\tilde{\mathcal{D}}_{1,k}^{J}
    +\frac{1+2x_{1}}{2+2x_{1}}\left[\mathcal{A}_{1,k}^{J\setminus j }-\frac{1}{2+2x_{1}}\tilde{\mathcal{D}}_{1,k}^{J\setminus  j }\right]\tilde{\mathcal{D}}_{1,j}^{J }\,,
\end{equation}
or alternatively  
\begin{equation}\label{eq:Coeff_c_2_S}
\mathcal{C}_{1,j,k}^{J}  =\frac{1+2x_{1}}{2x_{1}} \left[ \frac{1}{2+2x_{1}}\mathcal{A}_{1,j}^{J\setminus k }+\tilde{\mathcal{D}}_{1,j}^{J\setminus  k }\right]{\mathcal{A}}_{1,k}^{J}
    +\frac{1+2x_{1}}{2x_{1}}\left[\frac{1}{2x_{1}}\mathcal{A}_{1,k}^{J\setminus  j }+\tilde{\mathcal{D}}_{1,k}^{J\setminus  j}\right]{\mathcal{A}}_{1,j}^{J }\,.
\end{equation}


%
 
%

\bibliographystyle{utphys2}
\bibliography{qref}

\end{document}